\documentclass[conference]{IEEEtran}
\ifCLASSINFOpdf
\else
\fi
\usepackage{amsmath,dsfont,bbm,epsfig,amssymb,amsfonts,amstext,verbatim,amsopn,cite,subfigure,multirow,multicol,lipsum,xfrac}
\usepackage{amsthm}
\usepackage{mathtools,amsthm}
\usepackage{perpage}
\usepackage{url}
\usepackage{amsfonts}
\usepackage{epsfig}
\usepackage[font=small]{caption}
\usepackage{psfrag}	
\usepackage[Algorithm,ruled]{algorithm}
\usepackage{algorithmicx}
\usepackage{algpseudocode}
\usepackage{pifont}
\usepackage[utf8]{inputenc}
\usepackage[T1]{fontenc}  
\usepackage[nolist]{acronym}
\MakePerPage{footnote}
\usepackage{paralist}
\usepackage{enumitem}
\usepackage{bbm}
\usepackage[process=auto]{pstool}
\usepackage{tikz}
\usetikzlibrary{shapes,arrows}

\newcommand{\setX}{\mathbbmss{X}}
\newcommand{\setA}{\mathbbmss{A}}

\newcommand{\setR}{\mathbbmss{R}}

\newcommand{\setS}{\mathbbmss{S}}

\newcommand{\setZ}{\mathbbmss{Z}}

\newcommand{\setC}{\mathbbmss{C}}

\newcommand{\rmP}{P_{\max}}

\newcommand{\rmp}{\mathrm{p}}

\newcommand{\rmR}{\mathrm{R}}

\newcommand{\rmG}{\mathrm{G}}
\newcommand{\rmQ}{\mathrm{Q}}

\newcommand{\out}{\mathrm{out}}
\newcommand{\In}{\mathrm{in}}

\newcommand{\her}{\mathsf{H}}

\newcommand{\argmin}{\mathop{\mathrm{argmin}}}
\newcommand{\argmax}{\mathop{\mathrm{argmax}}}

\newcommand{\sfD}{\mathsf{D}}

\newcommand{\sfp}{\mathsf{p}}

\newcommand{\mae}{\mathcal{E}}

\newcommand{\mao}{\mathcal{O}}

\newcommand{\bxx}{\mathbf{x}}
\newcommand{\bhxx}{\mathbf{x}}
\newcommand{\bss}{\mathbf{s}}
\newcommand{\bww}{\mathbf{w}}
\newcommand{\buu}{\mathbf{u}}

\newcommand{\byy}{\mathbf{y}}
\newcommand{\bgg}{\mathbf{g}}
\newcommand{\bff}{\mathbf{f}}
\newcommand{\bhzz}{{\mathbf{z}}}

\newcommand{\rmx}{\mathrm{x}}
\newcommand{\rms}{\mathrm{s}}
\newcommand{\rmw}{\mathrm{w}}
\newcommand{\rmu}{\mathrm{u}}

\newcommand{\rmy}{\mathrm{y}}
\newcommand{\papr}{\mathrm{PAPR}}

\newcommand{\bx}{{\boldsymbol{x}}}

\newcommand{\xx}{\mathrm{x}}

\newcommand{\set}[1]{\left\lbrace#1\right\rbrace}

\newcommand{\bz}{{\boldsymbol{z}}}
\newcommand{\ba}{{\boldsymbol{a}}}

\newcommand{\bs}{{\boldsymbol{s}}}

\newcommand{\bv}{{\boldsymbol{v}}}

\newcommand{\bzz}{{\mathbf{z}}}

\newcommand{\dif}{\mathrm{d}}

\newcommand{\trp}{\mathsf{T}}

\newcommand{\mF}{\mathbf{F}}
\newcommand{\mR}{\mathbf{R}}

\newcommand{\mI}{\mathbf{I}}

\newcommand{\mJ}{\mathbf{J}}
\newcommand{\mG}{\mathbf{G}}
\newcommand{\mQ}{\mathbf{Q}}

\newcommand{\mU}{\mathbf{U}}
\newcommand{\mGam}{\mathbf{A}}
\newcommand{\mD}{\mathbf{D}}

\newcommand{\mV}{\mathbf{R}}

\newcommand{\mH}{\mathbf{H}}

\newcommand{\E}{\mathbb{E}}

\newcommand{\norm}[1]{\lVert #1 \rVert}
\newcommand{\re}[1]{\mathsf{Re}\left\lbrace #1 \right\rbrace}
\newcommand{\img}[1]{\mathsf{Im}\left\lbrace #1 \right\rbrace}

\newcommand{\abs}[1]{\lvert #1 \rvert}
\newcommand{\tr}[1]{\mathrm{Tr} \{ #1 \}}

\newcommand{\pn}[1]{( #1 )}
\newcommand{\glse}[1]{\mathrm{glse}\left( #1 \right)}
\newcommand{\gred}[1]{\nabla_{\hspace*{-.5mm}#1} \hspace*{.5mm}}
\newcommand{\gredSq}[1]{\nabla^2_{\hspace*{-.5mm}#1} \hspace*{.5mm}}

\newtheoremstyle{mystyle}
  {}
  {}
  {}
  {}
  {\bfseries}
  {:}
  { }
  {}

\theoremstyle{mystyle}

\newtheorem{remark}{Remark}

\newtheorem*{tun}{Tuning Strategy}
\newtheorem*{derivation}{Derivation}
\algnewcommand\algorithmicLet{\textbf{Let}}
\algnewcommand\Let{\item[\algorithmicLet]}
\algnewcommand\algorithmicSet{\textbf{Set}}
\algnewcommand\Set{\item[\algorithmicSet]}

\algnewcommand\algorithmicInitiate{\textbf{Initiate}}
\algnewcommand\Initiate{\item[\algorithmicInitiate]}
\algnewcommand\algorithmicStart{\textbf{Begin}}
\algnewcommand\Begin{\item[\algorithmicStart]}
\algnewcommand\algorithmicEnd{\textbf{End}}
\algnewcommand\End{\item[\algorithmicEnd]}

\algnewcommand\algorithmicOutP{\textbf{Output:}}
\algnewcommand\Out{\item[\algorithmicOutP]}
\newcommand\NoDo{\renewcommand\algorithmicdo{}}

\newcounter{bar}

\begin{document}

\begin{acronym}
\acro{mimo}[MIMO]{Multiple-Input Multiple-Output}
\acro{csi}[CSI]{Channel State Information}
\acro{awgn}[AWGN]{Additive White Gaussian Noise}
\acro{iid}[i.i.d.]{independent and identically distributed}
\acro{ut}[UT]{User Terminal}
\acro{bs}[BS]{Base Station}
\acro{tas}[TAS]{Transmit Antenna Selection}
\acro{glse}[GLSE]{Generalized Least Square Error}
\acro{rhs}[r.h.s.]{right hand side}
\acro{lhs}[l.h.s.]{left hand side}
\acro{wrt}[w.r.t.]{with respect to}
\acro{rs}[RS]{Replica Symmetry}
\acro{rsb}[RSB]{Replica Symmetry Breaking}
\acro{papr}[PAPR]{Peak-to-Average Power Ratio}
\acro{rzf}[RZF]{Regularized Zero Forcing}
\acro{snr}[SNR]{Signal-to-Noise Ratio}
\acro{rf}[RF]{Radio Frequency}
\acro{mf}[MF]{Match Filtering}
\acro{gamp}[GAMP]{Generalized Approximate Message Passing}
\acro{vamp}[VAMP]{Vector Approximate Message Passing}
\acro{map}[MAP]{Maximum-A-Posterior}
\acro{mmse}[MMSE]{Minimum Mean-Square-Error}
\acro{ap}[AP]{Average Power}
\end{acronym}

\title{Precoding via Approximate Message Passing with Instantaneous Signal Constraints\vspace*{-3mm}}
\author{
\IEEEauthorblockN{
Ali Bereyhi\IEEEauthorrefmark{1},
Mohammad Ali Sedaghat\IEEEauthorrefmark{2},
Ralf R. M\"uller\IEEEauthorrefmark{1},
}
\IEEEauthorblockA{
\IEEEauthorrefmark{1}Friedrich-Alexander Universit\"at Erlangen-N\"urnberg (FAU), \IEEEauthorrefmark{2}Cisco Optical GmbH N\"urnberg\\
ali.bereyhi@fau.de, mohammad.sedaghat@fau.de, ralf.r.mueller@fau.de\vspace*{-5mm}
\thanks{This work was supported by the German Research Foundation, Deutsche Forschungsgemeinschaft (DFG), under Grant No. MU 3735/2-1.}
}
}


\IEEEoverridecommandlockouts

\maketitle

\begin{abstract}
This paper proposes a low complexity precoding alg- orithm based on the recently proposed Generalized Least Square Error (GLSE) scheme with generic penalty and support. The algorithm iteratively constructs the transmit vector via Approximate Message Passing (AMP). Using the asymptotic decoupling property of GLSE precoders, we derive closed form fixed point equations to tune the parameters in the proposed algorithm for a general set of instantaneous signal constraints. The~tuning~strategy is then utilized to construct transmit vectors with restricted peak-to-average power ratios and to efficiently select a subset of transmit antennas. The numerical investigations show that the proposed algorithm tracks the large-system performance of GLSE precoders even for~a~moderate~number~of~antennas.
\end{abstract}

\IEEEpeerreviewmaketitle

\section{Introduction}
\label{sec:intro}


For a given precoding support $\setX\subset\setC$ and penalty function $u(\cdot):\setX\mapsto\setR$, the \ac{glse} precoder constructs the transmit vector $\bx\in\setX^N$ from the data vector $\bs\in \setC^K$ and the channel matrix $\mH\in\setC^{N\times K}$ as $\bx=\glse{\bs,\rho|\mH}$ where $\rho$ is a power control factor and \cite{bereyhi2017wsa}
\begin{align}
\glse{\bs,\rho|\mH}=\argmin_{\bv \in {\setX^N}} \norm{\mH \bv-\sqrt{\rho} \hspace*{.5mm}\bs}^2 +u(\bv). \label{eq:glse}
\end{align}
The generality of $\setX$ and $u(\cdot)$ allows for addressing various forms of constraints on the transmit vector. Compared to the classical approaches for imposing such constraints, the studies in \cite{sedaghat2016lse,sedaghat2017newclass,bereyhi2017wsa,bereyhi2017isit} have shown significant enhancements obtained via the \ac{glse} precoding scheme. Nevertheless, the computational complexity of this scheme has been remained as the main chal- lenge and is intended to be addressed in this paper. 

The main motivation of this study comes from the great deal of interest being received recently by massive \ac{mimo} systems \cite{hoydis2013massive}. Form implementational points of view, however, these systems confront the problem of high \ac{rf}-cost which raises due to the vast number of \ac{rf}-chains needed in such setups. The initial approach to overcome this issue is to restrict the \ac{papr} of the transmit vector \cite{mohammed2013per,chen2017low}. In this case, nonlinear power amplifiers with lower dynamic ranges can be employed, and the total \ac{rf}-cost can be significantly reduced. Another approach is \ac{tas} \cite{li2014energy,asaad2017tas} in which a subset of transmit antennas is kept active at each transmission interval, and therefore, the number of required \ac{rf}-chains is reduced. Although such approaches combat the issue of high \ac{rf}-cost, the conventional algorithms significantly degrade the performance. In this case, \ac{glse} precoders reduce this degradation by finding the optimal transmit vector which satisfies the constraints imposed by these approaches.
In general, \ac{glse} precoders solve an optimization problem in each transmission interval. This task is not trivial for choices of $u(\cdot)$ and $\setX$ which are non-convex. For cases with convex optimization problems, the precoder can be implemented via generic linear programming algorithms. The high computational complexity of these algorithms for large dimensions, however, leaves the implementation of \ac{glse} precoders as an issue in massive \ac{mimo} setups. \ac{gamp} \cite{rangan2011generalized} proposes a low complexity iterative approach for several estimation problems based on approximating the loopy belief propagation algorithm in the large limit \cite{donoho2009message}. The algorithm is known to considerably outperform other available iterative approaches. The underlying estimation problems, which are addressed by \ac{gamp}, are mathematically similar to the \ac{glse} precoding scheme, and therefore, the algorithm can be employed to design a class of iterative precoders based on the \ac{glse} scheme.

The main contribution of this paper is to adopt and tune the \ac{gamp} algorithm to address the \ac{glse} precoding scheme, recently proposed in \cite{sedaghat2016lse,sedaghat2017newclass,bereyhi2017wsa,bereyhi2017isit}. The developed iterative scheme is referred to as ``\ac{glse}-\ac{gamp}'' precoding and exhibits low complexity characteristic. Using the fact that the \ac{glse} and \ac{glse}-\ac{gamp} precoders consider same optimization problems, we further propose a tuning strategy based~on~the~asymptotic results in \cite{sedaghat2016lse,sedaghat2017newclass,bereyhi2017wsa,bereyhi2017isit} derived via the replica method. Our numerical investigations show that the performance of \ac{glse}-\ac{gamp} precoders tuned by the proposed strategy is accurately consistent with asymptotics of corresponding \ac{glse} precoders.

%
%

\subsection*{Notation}
Throughout the paper, scalars, vectors and matrices are represented with non-bold, bold lower case and bold upper case letters, respectively. $\mI_K$ is a $K \times K$ identity matrix, and $\mH^{\her}$ is the Hermitian of $\mH$. The set of real and integer numbers are denoted by $\setR$ and $\setZ$, and $\setC$ represents the complex plane. For $s\in\setC$, $\re{s}$, $\img{s}$ and $\bss\coloneqq \left[ \re{s} \hspace*{1mm} \img{s} \right]^\trp$ identify the real part, imaginary part and augmented vector, respectively, 
and the expression $\bss \in \setS$ indicates that $\bss$ is the augmented version of $s\in\setS$. For $\bff(\bx)=\left[ f_1(\bx), \ldots, f_n(\bx) \right]^\trp$,
the gradient operator is defined as $\gred{\bx} \bff(\bx) \coloneqq [ \gred{\bx} f_1(\bx), \ldots,\gred{\bx} f_n(\bx) ]^\trp$. $\norm{\cdot}$ and $\norm{\cdot}_1$ denote the Euclidean and $\ell_1$-norm, respectively. Considering the random variable $x$, $\mathrm{p}_x$ represents either the probability mass or density function. Moreover, $\hspace*{-.1mm}\E\hspace*{-.1mm}$~identifies~the expectation. For sake of compactness, $\set{1, \ldots , N}$ is abbreviated by $[N]$, and we define $\tilde{\phi}(x,\lambda)\coloneqq \exp(-{x^2}/{\lambda})$ and 
$\tilde{\rmQ}(x,\lambda) \coloneqq \int_x^\infty \tilde{\phi}(u,\lambda) \dif u / \lambda$
for a given non-negative real $\lambda$.

\section{Problem Formulation}
\label{sec:sys}
Consider a Gaussian broadcast \ac{mimo} setup in which a sequence of data symbols $\set{s_k}$ for $k\in [K]$ is transmitted to $K$ single-antenna users simultaneously. The transmitter is equipped with $N$ transmit antennas. The channel is considered to be quasi-static fading and perfectly known at the transmitter. By employing the \ac{glse} precoding scheme given in \eqref{eq:glse} with some penalty $u(\cdot)$ and precoding support $\setX\subseteq \setC$, the transmit vector is constructed as $\bx_{N\times 1}=\glse{\bs,\rho|\mH}$~where~$\bs_{K\times 1}\coloneqq \left[ s_1, \ldots, s_K\right]^\trp$ 
and $\rho$ is a non-negative power control factor. For this setup, we assume that the following constraints hold.
\begin{enumerate}[label=(\alph*)]
\item $\bs_{K \times 1}$ has \ac{iid} zero-mean complex Gaussian entries with unit variance.
\item $u(\cdot)$ decouples meaning that $u(\bv)=\sum_{j=1}^N u(v_j)$.
\item $N$ and $K$ grow large, such that the load factor $\alpha\coloneqq K/N$ is kept fixed in both $N$ and $K$.
\item $\mH^{\her} \mH = \mU \mD \mU^{\her}$ in which $\mU$ is an $N\times N$ unitary matrix, and $\mD$ is a diagonal matrix with asymptotic eigenvalue distribution $\rmp_\mD$. For $\rmp_\mD$, we define the Stieltjes transform as $\rmG_\mD(s)= \E \left\lbrace (d-s)^{-1} \right\rbrace $ with the expectation being taken over $d\sim \rmp_\mD$ and the $\rmR$-transform as $\rmR_\mD (\omega) = \rmG_\mD^{-1} (-\omega)- \omega^{-1}$ where $\rmG_\mD^{-1} (\cdot)$ denotes the inverse with respect to composition.
\end{enumerate}
By proper choices of the support $\setX$ and penalty $u(\cdot)$, the \ac{glse} precoder can impose several constraints on the transmit vector. 
\begin{itemize}
\item Setting $u(\bv)=\lambda\norm{\bv}^2$ and $\setX=\set{x \in \setC: \abs{x}^2<P}$, the transmit vector is restricted to have a limited \ac{papr}. In fact in this case, the peak power is set to $P$ and a desired constraint on the \ac{papr} is imposed by tuning $\lambda$ such that the average power is accordingly restricted.
\item Let $u(\bv)\hspace*{-.5mm}=\hspace*{-.5mm}\lambda\norm{\bv}^2+\mu\norm{\bv}_1$ and $\setX = \setC$; then, the number of active transmit antennas is constrained.
\end{itemize}

\section{\ac{glse}-\ac{gamp} Precoders}
\label{sec:result}
The \ac{glse} scheme can be considered as a max-sum problem which can be addressed via the \ac{gamp} algorithm~\cite{rangan2011generalized}. 
\subsection{\ac{gamp} Algorithm}
The \ac{gamp} algorithm, proposed in \cite{rangan2011generalized}, intends to estimate $\bv_{N\times 1}$ from $\bs_{K\times 1}$ iteratively considering the following setup.
\begin{enumerate}[label=(\alph*)]
\item Each entry of $\bv$ is generated from the corresponding entry of some $\ba\in\setA^{N}$ via $\rmp_{v|a}$.
\item The entries of $\bs$ are obtained form the entries of the vector $\bz_{K\times 1}$ through identical scalar channels with $\rmp_{s|z}$.
\item $\bz$ is a random linear transform of $\bv$, i.e., $\bz=\mH \bv$ for some random $K\times N$ matrix $\mH$.
\end{enumerate}
Depending on the estimation scheme, the \ac{gamp} algorithm~is developed to address the ``max-sum'' or ``sum-product'' problems. The max-sum \ac{gamp} algorithm iteratively determines the \ac{map} estimation 
\begin{align}
\bx = \argmax_{\bv} \sum_{n=1}^N f_\In(v_n,a_n)+\sum_{k=1}^K f_\out(z_k,y_k) \label{eq:map}
\end{align}
for some scalar functions $f_\In(\cdot,\cdot)$ and $f_\out(\cdot,\cdot)$ which represent the conditional distributions $\rmp_{v|a}$ and $\rmp_{s|z}$. The sum-product \ac{gamp} algorithm, moreover, addresses the \ac{mmse} estimation where $\bx = \E\set{\bv|\bs,\ba}$.
\subsection{The \ac{gamp}-\ac{glse} Algorithm}
By comparing \ac{glse} precoding with \eqref{eq:map}, it is observed that the precoding scheme solves a max-sum problem in which $\bz\coloneqq\mH \bv$ with $\mH$ being the channel matrix, $v_k \in \setX$ for $k\in [K]$, and 
$f_\In(v_n,a_n) = - u(v_n)$ and 
$f_\out(z_k,s_k) = - \abs{z_k-\sqrt{\rho} s_k}^2$. 
As the result, the \ac{gamp} algorithm can be applied to iteratively construct the transmit vector $\bx$. By some lines of derivations, the max-sum \ac{gamp} algorithm can be adopted to the \ac{glse} scheme in \eqref{eq:glse}. The resulting algorithm is referred to as ``\ac{glse}-\ac{gamp}'' algorithm and is represented in Algorithm~\ref{A-GAMP} for the precoding support $\setX \subseteq \setC$ and the complex-valued matrix $\mH$. The variables and functions in the algorithm, for $k\in [K]$ and $n\in [N]$, are defined as follows.
\begin{itemize}
\item The real two-dimensional vectors $\bww_k$, $\bzz_k$, $\byy_k$, $\bss_k$, $\buu_n$ and $\bxx_n$ are the augmented forms of the complex scalars $w_k$, $z_k$, $y_k$, $s_k$, $u_n$ and $x_n$, respectively. 
\item The matrices $\mR^\rmw_k$, $\mR^\rmy_k$, $\mR^\rmu_n$ and $\mR^\rmx_n$ are real $2\times 2$ matrices, and $\mQ_{kn}$ is defined as
\begin{align}
\mQ_{kn} \coloneqq 
\begin{bmatrix}
\re{h_{kn}} &-\img{h_{kn}}\\
\img{h_{kn}} &\hphantom{-}\re{h_{kn}}
\end{bmatrix}
\end{align}
with $h_{kn}$ representing the entry $(k,n)$ of $\mH$.\vspace*{1mm}
\item $\bgg_{\out}\left(\cdot \right)$ is the output thresholding function defined as
\begin{align}
\bgg_{\out}\left(\bww,\bss,\mV\right) &\coloneqq \gred{\bww} \min_{\bzz\in\setC} \mae_{\out} \pn{\bzz,\bww,\bss,\mV} \label{eq:g_out}
\end{align}
where the function $\mae_{\out} \pn{\cdot}$ is determined by
\begin{align}
\mae_{\out} \pn{\bzz,\bww,\bss,\mV} = \frac{1}{2} &\pn{\bzz - \bww}^\trp  \mV^{-1} \pn{\bzz - \bww} \nonumber \\
 &+  \norm{\bzz-\sqrt{\rho} \hspace*{.5mm} \bss}^2
\end{align}
\item $\bgg_{\In}\left(\cdot \right)$ is the input thresholding function~being defined as
\begin{align}
\bgg_{\In}\left( \buu,\mV\right) \coloneqq \argmin_{\bxx\in\setX} \mae_{\In} \pn{\bxx,\buu, \mV}. \label{eq:g_In}
\end{align}
where the function $\mae_{\In} \pn{\cdot}$ is evaluated by
\begin{align}
\mae_{\In} \pn{\bxx,\buu,\mV} =  \frac{1}{2} \pn{\buu-\bxx}^\trp \mV^{-1} \pn{\buu-\bxx} + u(\bxx). \label{eq:E_in}
\end{align}
\item The initial conditions are 
$\bxx_n\pn{1}= \arg \min_{\bxx\in\setX} u(\bxx)$ and 
$\mV_n^{\xx}\pn{1}= \left[\gredSq{\bxx} u\pn{\buu_n\pn{1}}\right]^{-1}$. 
\end{itemize}

The update rules in Algorithm~\ref{A-GAMP} are derived by extending the sum-max \ac{gamp} algorithm to the case with a complex-valued matrix $\mH$ and an arbitrary input support $\setX\subseteq \setC$. The extension is followed by determining the update rules for the corresponding loopy belief propagation algorithm and then taking some steps similar to \cite[Appendix~C]{rangan2011generalized}. The detailed derivations are skipped due to the page limit and is represented in the extended version of the manuscript. 
\begin{remark}
One should distinguish between the  \ac{glse} scheme and the \ac{glse}-\ac{gamp} algorithm. In fact, the former is a least square based scheme to design transmit signals which fulfill some desired constraints. The \ac{glse}-\ac{gamp} algorithm, on the other hand, proposes an iterative approach based on \ac{gamp} to address the \ac{glse} scheme. For some choices of the penalty function, precoding support and channel matrix, the \ac{glse}-\ac{gamp} algorithm converges to the transmit signal given by the \ac{glse} scheme. There are however some particular cases in which the \ac{glse}-\ac{gamp} algorithm does not converge. For these cases, Algorithm~\ref{A-GAMP} does not give the desired transmit signal. To avoid the divergence in such cases, we need to modify the algorithm. This issue is briefly discussed in Section~\ref{sec:num}.  
\end{remark}
In contrast to \ac{glse} precoders, \ac{glse}-\ac{gamp} precoders exhibit low complexity characteristic. Considering Algorithm~\ref{A-GAMP} and noting that the matrices in \eqref{eq:A1}-\eqref{eq:A9} are fixed $2\times 2$ matrices, it is straightforward to show that the total worst-case complexity of \ac{glse}-\ac{gamp} precoders per iteration is $\mao(KN)$. The number of iterations, moreover, does not grow with the dimensions. Therefore, one can conclude that the overall complexity of the precoding scheme is $\mao(KN)$~as~well. 

\begin{figure}
\vspace*{-2.5mm}
\end{figure}
\begin{algorithm}[t]
\caption{ \ac{glse}-\ac{gamp} Precoding Algorithm}
\label{A-GAMP}
\begin{algorithmic}[0]
\Initiate Start from $t=1$ and for $ k \in[K]$ let $\byy_k \pn{0} =\boldsymbol{0}$. Set $\bhxx_n\pn{1}$ and $\mV_n^{\rmx}\pn{1}$ for $ n\in[N]$ to their initial conditions. \vspace*{2mm}
\While\NoDo $t< T$
\For\NoDo $k\in[K]$
\begin{subequations}
\begin{align}
\mV_k^{\rmw} \pn{t} &= \sum_{n=1}^N \mQ_{k n} \mV_n^{\rmx} \pn{t} \mQ_{k n}^\trp \label{eq:A1} \\
\bhzz_k \pn{t} &=\sum_{n=1}^N \mQ_{k n} \bhxx_n \pn{t} \label{eq:A2} \\
\bww_k \pn{t} &=\bhzz_k\pn{t} -\mV_k^{\rmw} \pn{t} \byy_k \pn{t-1} \label{eq:A3} \\
\byy_k \pn{t} &=\bgg_{\out}( \bww_k \pn{t} ,\bss_k,\mV_k^{\rmw} \pn{t} ) \label{eq:A4} \\
\mV_k^{\rmy} \pn{t} &=-\gred{\bww} \bgg_{\out} ( \bww_k \pn{t} ,\bss_k,\mV_k^{\rmw} \pn{t} )\label{eq:A5}
\end{align}
\end{subequations}
\EndFor
\For\NoDo $n \in[N]$
\begin{subequations}
\begin{align}
\mV_n^{\rmu} \pn{t} &= \left[ \sum_{k=1}^K \mQ_{k n}^\trp \mV_{k}^{\rmy} \pn{t} \mQ_{k n}\right]^{-1} \label{eq:A6} \\
\buu_n \pn{t} &=\bhxx_n \pn{t} + \mV_n^{\rmu} \pn{t}\left[ \sum_{k=1}^K \mQ_{k n}^\trp \byy_k \pn{t} \right] \label{eq:A7} \\
\bhxx_n \pn{t+1} &=\bgg_{\In} ( \buu_n \pn{t} ,\mV_n^{\rmu} \pn{t} ) \label{eq:A8} \\
\mV_n^{\rmx} \pn{t+1} &= \left[\gred{\buu} \bgg_{\In} ( \buu_n \pn{t} ,\mV_n^{\rmu} \pn{t} ) \right] \mV_n^{\rmu} \pn{t} \label{eq:A9}
\end{align}
\end{subequations}
\EndFor
\EndWhile \vspace*{1mm}
\Out $\bxx_n (T)$ for $n \in [N]$.
\end{algorithmic}
\end{algorithm}

\subsection{Tuning \ac{glse}-\ac{gamp} precoders}
\label{sec:tune_gen}
In order to impose a given set of constraint on the transmit signal, the corresponding \ac{glse}-\ac{gamp} precoder should be tuned. As an example, consider the case in which the number of active transmit antennas, as well as the average transmit power, is desired to be restricted via a \ac{glse}-\ac{gamp} precoder. In this case, one may set $\setX = \setC$ and $u(\bv)=\lambda\norm{\bv}^2+\mu\norm{\bv}_1$. The factors $\lambda$ and $\mu$ in this case control the average transmit power and the fraction of active antennas, respectively. Consequently for given constraints, these factors need to be tuned. Nevertheless, the derivation of an exact tuning strategy is not a trivial problem as the constrained parameters, i.e.,~the~ave- rage power or fraction of active antennas, cannot be derived in terms of the tuning factors straightforwardly. We therefore propose a tuning strategy based on the asymptotics of the \ac{glse}-\ac{gamp} algorithm and its connection to the \ac{glse} scheme. The large-system performance of \ac{glse}-\ac{gamp} precoders is studied through asymptotic analyses of ``state evolution'' equations; see \cite{javanmard2013state} and the references therein. Following the results in the literature, e.g. \cite{rangan2014convergence,rangan2016fixed}, it is shown that for choices of $\mH$, $\setX$ and $u(\cdot)$, in which the \ac{glse}-\ac{gamp} algorithm converges, the asymptotic performance of the algorithm coincides with the large-system performance of \ac{glse} precoders investigated in \cite{bereyhi2017wsa,bereyhi2017isit}. This result indicates that in the large-system limit, the tuning factors for \ac{glse}-\ac{gamp} and \ac{glse} precoders are the same. Therefore, for a given set of constraint, we derive the tuning factors of the \ac{glse}-\ac{gamp} precoders by tuning the corresponding \ac{glse} precoders. 
\begin{tun}
Assume that the constraints $f_j(\bx)/N=C_j$ are desired to be satisfied via a \ac{glse}-\ac{gamp} precoder with penalty $u(\cdot)$ and support $\setX$ which are controlled by $\lambda_j$~for~$j\in [J]$. Here, $f_j(\cdot)$ are decoupling functions meaning that $f_j(\bx)=\sum_{n=1}^N f_j(x_n)$. To tune $\lambda_j$ accordingly, we define 
\begin{align}
\xx=\argmin_{v\in\setX} \abs{v- s_0}^2+ \xi \hspace*{1mm} u(v) \label{eq:snigle}
\end{align}
where $s_0\sim \mathcal{CN}\left( 0, \sigma^2 \right)$ with
\begin{align}
\sigma^2=\left[\rmR_\mD(-\chi)\right]^{-2}\frac{\partial}{\partial \chi} \left[ ( \lambda_s \chi- \sfp ) \rmR_\mD(-\chi) \right].
\end{align}
and $\xi=\left[\rmR_\mD(-\chi)\right]^{-1}$ for $\chi$ and $\sfp$ which satisfy  $\sfp= \E \abs{\xx}^2$~and
\begin{align}
 \frac{\sigma^2 \chi}{\xi} &=  \E \re{\xx^* s_0} . \label{eq:fix}
\end{align}
The precoder is then accordingly tuned by choosing $\lambda_j$ for $j\in [J]$ such that the equations  $\E f_j(\xx)=C_j$ are satisfied.
\end{tun}
\begin{derivation}
The derivation follows~the~marginal decoupling property of the \ac{glse} precoders presented in \cite{bereyhi2017wsa,bereyhi2017isit}. In fact, using the property, it is concluded that $f_j(\bx)/N$ asymptotically converges to $\E f_j(\xx)$. By taking the approach illustrated at the beginning of the section, the tuning strategy is obtained.
\end{derivation}
The proposed tuning strategy evaluate the decoupled \ac{glse} precoder\footnote{See Proposition~2 in \cite{bereyhi2017wsa} for the decoupling property of \ac{glse} precoders. A more general version of the property is represented in \cite[Section II-A]{bereyhi2017isit}.} by finding $\chi$ and $\sfp$ form the fixed-point equations. The asymptotic constrained parameters are then determined by taking the expectation $\E f_j(\xx)$ and set it equal to $C_j$. One should note that the strategy in general is heuristic, since it tunes the precoders for the large-system limit. Nevertheless, the numerical investigations show that for several cases, the \ac{glse}-\ac{gamp} precoders are well tuned via this strategy.

\section{Applications of \ac{glse}-\ac{gamp} Precoders}
In this section, we investigate two special cases of \ac{glse}-\ac{gamp} precoders with \ac{tas} and limited \ac{papr}. Throughout the analyses, we assume that $\mH$ represents an \ac{iid} Rayleigh fading channel with variance $1/N$, i.e., $\rmR_{\mJ}(\omega)=\alpha (1-\omega)^{-1}$. 

\subsection{\ac{glse}-\ac{gamp} Precoder with \ac{tas}}
\label{sec:tas-gamp}

As it was discussed, \ac{tas} can be directly addressed at the transmit side by using \ac{glse} scheme with $u(v)=\lambda \abs{v}^2 + \mu \abs{v}$. 
The corresponding \ac{glse}-\ac{gamp} precoder is therefore given by Algorithm~\ref{A-GAMP} 
where $\setX=\setC$, 
$\bgg_{\out}\left(\bww,\bss,\mV\right) = \mG_{\rmw} \bww + \mG_{\rms} \bss$ and 
$\gred{\bww}\bgg_{\out}\left(\bww,\bss,\mV\right) = \mG_{\rmw}$, respectively 
with 
\begin{subequations}
\begin{align}
\mG_{\rmw} &\coloneqq-2\hspace*{.5mm}\mGam^{\trp}\mGam- \pn{\mGam-\mI_2}^\trp \mV^{-1} \pn{\mGam-\mI_2}  \\
\mG_{\rms} \hspace*{1mm} &\coloneqq-2\left[ 2\mGam^{\trp}\mGam\mV-\mGam^{\trp}+\pn{\mGam-\mI_2}^\trp \mV^{-1} \mGam \mV\right]
\end{align}
\end{subequations}
and $\mGam\coloneqq\pn{\mI_2+2\mV}^{-1}$. For the input thresholding function, the analytic evaluation of the function from the augmented form in \eqref{eq:g_In} is not trivial. We thus employ the complex scalar form of the equation which results in $\bgg_{\In}\left( \buu,\mV\right) = \mG (\buu) \bff(\buu)$ and $\gred{\buu} \bgg_{\In}\left( \buu,\mV\right) = \mG (\buu) \mF(\buu)$ where
\begin{align}
\mG (\buu) =
\begin{cases}
    \mG_{\rmu} \hspace*{.3mm}  \vspace*{.8mm}  \qquad & {\norm{\buu}} \geq \tau \\
    0             &  {\norm{\buu}} < \tau \label{eq:ave_in}
\end{cases}
\end{align}
with $\tau \coloneqq 2 \mu \left[\tr{\mV^{-1}}\right]^{-1}$ and $\mG_{\rmu} \coloneqq \pn{\mI_2+2\lambda\mV}^{-1}$, and
\begin{subequations}
\begin{align}
\bff(\buu) &\coloneqq \left[1-\frac{\tau}{\norm{\buu}}\right] \buu, \label{eq:bf_u}\\
\mF(\buu) &\coloneqq \frac{\tau}{\norm{\buu}^3} \buu\buu^\trp + \left[1-\frac{\tau}{\norm{\buu}}\right] \mI_2 . \label{eq:F_u}
\end{align}
\end{subequations}
By setting $\mu=0$, the \ac{glse} scheme reduces to \ac{rzf} precoding, and thus, the \ac{glse}-\ac{gamp} algorithm iteratively constructs the output of the \ac{rzf} precoder.
\subsubsection*{Tuning Strategy}
We employ the strategy in Section~\ref{sec:tune_gen} to tune $\mu$ and $\lambda$ such that the fraction of active antennas and the average transmit power are $\eta$ and $P$, respectively. For this case, $J=2$ and $f_1(\bx)=\norm{\bx}^2$ and $f_2(\bx)=\norm{\bx}_0$. Consequently, $\lambda$ and $\mu$ are determined from the fixed-point equations $\tilde{\phi}(\xi \mu  ;\theta) = \eta$ for $\theta=(\rho + P)/{\alpha}$ and 
\begin{align}
(1+2\xi \lambda)^2 &=\frac{\theta}{P} \left[\eta- 2 \hspace*{.2mm} \xi \mu \hspace*{.3mm} \tilde{\rmQ} ({ \xi \mu};\theta ) \right]
\end{align}
and $\xi$ is determined in terms of $\lambda$ and $\mu$ through
\begin{align}
\alpha \xi &= \frac{1}{2} + \frac{\xi}{1+2\xi \lambda} \left[\eta-  \xi \mu \hspace*{.3mm} \tilde{\rmQ} ({ \xi \mu};\theta ) \right].
\end{align}


\subsection{\ac{glse}-\ac{gamp} Precoder with \ac{papr} Constraint}
\label{sec:papr-gamp}
The precoder in Section~\ref{sec:tas-gamp} can further take the \ac{papr} constraint into account by setting $\setX \hspace*{-.5mm} = \hspace*{-.5mm} \set{ x\in\setC:  \abs{x}^2 < \rmP }$. The support in this case imposes a peak power constraint on the transmit signal which along with the penalty function restricts both the \ac{papr} and the number of active antennas\footnote{See \cite[Section IV-B]{bereyhi2017wsa} for further illustrations.}. Considering Algorithm~\ref{A-GAMP}, the output function for this setup remains unchanged , and the input function reads
%
\begin{align}
\bgg_{\In}\left(\buu,\mV\right) =
\begin{cases}
  \dfrac{{\buu}}{{\norm{\buu}}} \sqrt{\rmP}  \vspace*{1mm}   & \tilde{\tau} \leq {\norm{\buu}} \\
    \mG_{\rmu} \hspace*{.3mm} \bff(\buu) \vspace*{.8mm}  \qquad &\tau \leq {\norm{\buu}} < \tilde{\tau} \\
    0             & 0 \leq {\norm{\buu}} < \tau \label{eq:peak_in}
\end{cases}
\end{align}
with the corresponding gradient
\begin{align}
\gred{\buu}\bgg_{\In}\left(\buu,\mV\right) =
\begin{cases}
  \dfrac{\tilde{\buu}\tilde{\buu}^\trp}{{\norm{\buu}}^3} \sqrt{\rmP} \vspace*{.8mm}    & \tilde{\tau} \leq {\norm{\buu}} \\
    \mG_{\rmu} \hspace*{.3mm}\mF(\buu)\vspace*{.8mm}   \qquad &\tau \leq {\norm{\buu}} < \tilde{\tau} \\
    0             &0\leq {\norm{\buu}} < \tau, \label{eq:peak_grad_in}
\end{cases}
\end{align}
where $\tilde{\buu}\coloneqq \left[\rmu_2,-\rmu_1\right]^\trp$, $\tau \coloneqq 2 \mu \left[\tr{\mV^{-1}}\right]^{-1}$, and
\begin{align}
\tilde{\tau} \coloneqq \left(1+\frac{4\lambda}{\tr{\mV^{-1}}} \right) \sqrt{\rmP} + \frac{2\mu}{\tr{\mV^{-1}}}.
\end{align}
$\mG_\rmu$, $\bff(\buu)$ and $\mF(\buu)$ are moreover given as in Section~\ref{sec:tas-gamp}. 
By setting $\mu=0$, the~precoder~employs all the transmit antennas and restricts only the \ac{papr}. In this case, $\mF(\buu)=\mI_2$, $\bff(\buu)=\buu$, and $\tau$ reduces to zero.
\subsubsection*{Tuning Strategy}
Consider the same constraints as for the case without the \ac{papr} restriction. From Section~\ref{sec:tune_gen}, $\lambda$ and $\mu$ for the average power $P$ and the fraction of active antennas~$\eta$ are given by the fixed-point equations $\tilde{\phi}(\xi \mu  ;\theta) = \eta$ and 
\begin{align}
(1+2\xi \lambda)^2 &=\frac{\theta}{P} \left[\Delta_1 (\xi \mu) - 2 \hspace*{.2mm} \xi \mu \hspace*{.3mm} \Delta_2 (\xi \mu)\right].
\end{align}
Here, $\xi$ is a function of $\lambda$ and $\mu$ which satisfies
\begin{align}
\alpha \xi &= \frac{1}{2} + \frac{\xi}{1+2\xi \lambda} \left[\Delta_1 (\xi \mu) - 2 \hspace*{.2mm} \xi \mu \hspace*{.3mm} \Delta_2 (\xi \mu)\right].
\end{align}
Moreover, $\theta=(\rho + P)/{\alpha}$ and we have defined
\begin{subequations}
\begin{align}
\Delta_1(\xi \mu) &\coloneqq \tilde{\phi}(\xi \mu  ;\theta) - \tilde{\phi}(\xi \mu + (1+2\xi \lambda)\sqrt{\rmP}  ;\theta), \\
\Delta_2(\xi \mu) &\coloneqq \tilde{\rmQ} (\xi \mu  ;\theta) - \tilde{\rmQ} (\xi \mu + (1+2\xi \lambda)\sqrt{\rmP}  ;\theta) .
\end{align}
\end{subequations}
\section{Numerical Investigations}
\label{sec:num}
To investigate the performance of \ac{glse}-\ac{gamp} precoders, we define the distortion measure for a given $\rho$ as
\begin{align}
\sfD(\rho) \coloneqq \frac{1}{K} \E \norm{\mH\bx-\sqrt{\rho}\hspace*{.2mm} \bs}^2 \label{eq:D_rho}
\end{align}
which determines the average distortion caused by the multiuser interference at receive terminals. It is moreover shown that the achievable ergodic rate per user can be bounded from below in terms of $\sfD(\rho)$ as proved in \cite{sedaghat2016lse}.

\begin{figure}[t]
\hspace*{-.8cm}  
\resizebox{1.085\linewidth}{!}{
\pstool[width=.35\linewidth]{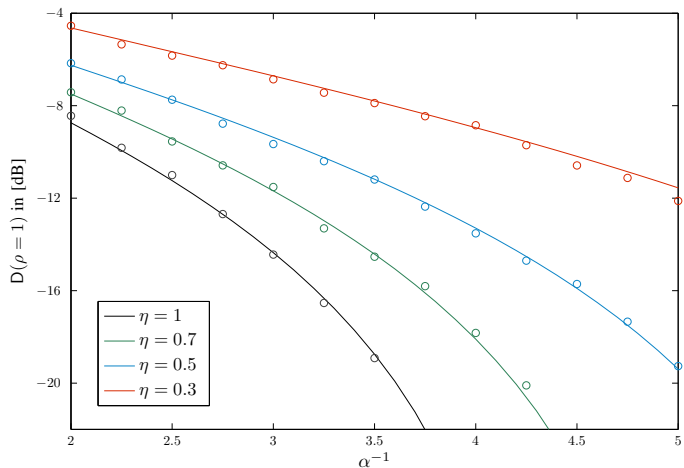}{
\psfrag{D}[c][t][0.23]{$\sfD(\rho=1)$ in [dB]}
\psfrag{alpha-inv}[c][c][0.25]{$\alpha^{-1}$}
\psfrag{PP=3-070}[l][l][0.24]{$\eta=0.7$}
\psfrag{PP=3-100}[l][l][0.24]{$\eta=1$}
\psfrag{PP=3-030}[l][l][0.24]{$\eta=0.3$}
\psfrag{PP=3-050}[l][l][0.24]{$\eta=0.5$}
\psfrag{-4}[r][c][0.18]{$-4$}
\psfrag{-8}[r][c][0.18]{$-8$}
\psfrag{-12}[r][c][0.18]{$-12$}
\psfrag{-16}[r][c][0.18]{$-16$}
\psfrag{-20}[r][c][0.18]{$-20$}

%
\psfrag{2}[c][b][0.18]{$2$}
\psfrag{2.5}[c][b][0.18]{$2.5$}
\psfrag{3}[c][b][0.18]{$3$}
\psfrag{3.5}[c][b][0.18]{$3.5$}
\psfrag{4}[c][b][0.18]{$4$}
\psfrag{4.5}[c][b][0.18]{$4.5$}
\psfrag{5}[c][b][0.18]{$5$}
}}
\caption{Distortion at $\rho=1$ vs. $\alpha^{-1}$ for $P=0.3$ and various $\eta$. Circles depict the performance of the \ac{glse}-\ac{gamp} precoder for $N=64$ and $T=20$. Solid lines denote the asymptotic performance of the corresponding \ac{glse} scheme determined by the replica method. 
\vspace*{-3.5mm}}
\label{fig:1}
\end{figure}

\begin{figure}[t]
\hspace*{-.8cm}  
\resizebox{1.085\linewidth}{!}{
\pstool[width=.35\linewidth]{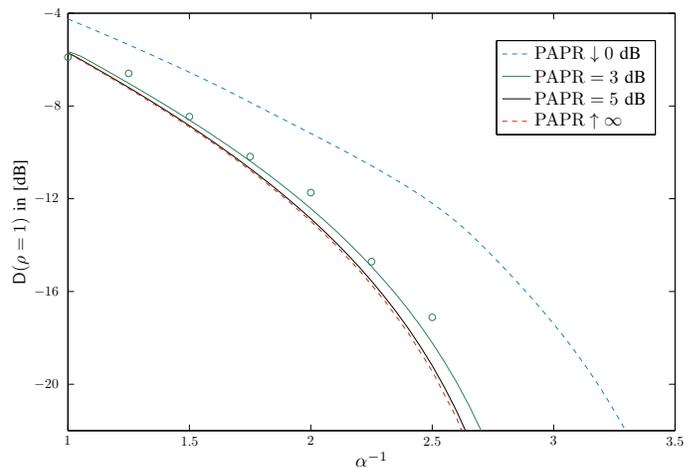}{
\psfrag{D}[c][t][0.23]{$\sfD(\rho=1)$ in [dB]}
\psfrag{alpha-inv}[c][c][0.25]{$\alpha^{-1}$}
\psfrag{PAPR=3-AAA-BBB}[l][l][0.24]{$\papr=3$ dB}
\psfrag{PAPR=0-AAA-BBB}[l][l][0.24]{$\papr \downarrow 0$ dB}
\psfrag{PAPR=5-AAA-BBB}[l][l][0.24]{$\papr=5$ dB}
\psfrag{PAPR=9-AAA-BBB}[l][l][0.24]{$\papr\uparrow\infty$}
\psfrag{-4}[r][c][0.18]{$-4$}
\psfrag{-8}[r][c][0.18]{$-8$}
\psfrag{-12}[r][c][0.18]{$-12$}
\psfrag{-16}[r][c][0.18]{$-16$}
\psfrag{-20}[r][c][0.18]{$-20$}

%
\psfrag{2}[c][b][0.18]{$2$}
\psfrag{2.5}[c][b][0.18]{$2.5$}
\psfrag{3}[c][b][0.18]{$3$}
\psfrag{3.5}[c][b][0.18]{$3.5$}
\psfrag{4}[c][b][0.18]{$4$}
\psfrag{1.5}[c][b][0.18]{$1.5$}
\psfrag{1}[c][b][0.18]{$1$}
}}
\caption{Distortion at $\rho=1$ vs. $\alpha^{-1}$ for several \ac{papr}s. $P=0.5$ and $\eta=1$. Solid lines and circles respectively denote the results for \ac{glse} and \ac{glse}-\ac{gamp} algorithm with $N=64$ and $T=20$. 
\vspace*{-4mm}}

\label{fig:2}
\end{figure}

The circles in Fig.~\ref{fig:1} show the distortion given by the \ac{glse}-\ac{gamp} precoder presented in Section~\ref{sec:tas-gamp} for various inverse load factors $\alpha^{-1}=N/K$ considering several constraints on the number of active antennas. The results have been given for $N=64$ antennas and $T=20$ iterations. The asymptotic performances of the corresponding \ac{glse} precoders, derived via the replica method in \cite{bereyhi2017isit}, have been also sketched with solid lines. 
Here, $\rho=1$ and $\lambda$ is set such that $P=0.3$. As the figure shows, the \ac{glse}-\ac{gamp} precoder tracks accurately the performance of the \ac{glse} scheme, even for a practically moderate number of antennas. 
For the \ac{papr}-limited precoder in Section~\ref{sec:papr-gamp}, the distortion at $\rho=1$ has been plotted in terms of $\alpha^{-1}$ in Fig.~\ref{fig:2}. The curves have been sketched for multiple \ac{papr} constraints. Similar to Fig.~\ref{fig:1}, solid lines correspond to the \ac{glse} scheme and circles denote the simulation results for the \ac{glse}-\ac{gamp} precoder with $N=64$ and $T=20$ for \ac{papr} $=3$ dB. Here, we have considered $P=0.5$, and $\rmP$ is tuned via the proposed strategy assuming all the antennas being active. 
The figure depicts that by increasing the \ac{papr} up to $5$ dB, the performance of the precoder is sufficiently close to the case without \ac{papr} restriction. This observation suggests for employing the \ac{glse}-\ac{gamp} precoder, in order to reduce the transmit \ac{papr} without any significant performance loss. In this case, low efficiency power amplifiers can be utilized which can significantly reduce the \ac{rf}-cost.

\begin{remark}
\label{remark:1}
It is known that the \ac{gamp} algorithm converges for \ac{iid} Gaussian matrices \cite{rangan2014convergence,rangan2016fixed}. However, by deviating from this assumption, the algorithm may diverge. This issue was recently addressed in \cite{rangan2017vector} via the \ac{vamp} algorithm. Consequently, for channel models with ill-conditioned matrices, one can develop a precoding algorithm based on the \ac{glse} scheme by taking a same approach while employing \ac{vamp}.
\end{remark}


%
%
\section{Conclusion}
\label{conclusion}
This paper has proposed a class of low complexity precoders based on the \ac{glse} scheme using the \ac{gamp} algorithm. 
The numerical investigations have been consistent with the replica results for the \ac{glse} scheme given in \cite{sedaghat2016lse,sedaghat2017newclass,bereyhi2017wsa,bereyhi2017isit}. This consistency demonstrates that various implementational limitations in massive \ac{mimo} systems can be effectively overcome using some low-complexity, but effective, algorithms. As indicated in Remark~\ref{remark:1}, the \ac{glse}-\ac{gamp} precoders may fail in converging for channel models with ill-conditioned channel matrices, and therefore, an alternative algorithm can be proposed via \ac{vamp}. The extension under \ac{vamp} is however skipped and left as a possible future work.

\bibliography{ref}
\bibliographystyle{IEEEtran}
\end{document}